\documentclass[%
 reprint,
superscriptaddress,
 amsmath,amssymb,
 aps,
]{revtex4-2}

\pdfoutput=1
\usepackage{graphicx}
\usepackage{dcolumn}
\usepackage{bm}
\usepackage{stackengine}
\usepackage{braket}
\usepackage{newtxmath}
\usepackage{hyperref}

\newcommand{\orcid}[1]{\href{https://orcid.org/#1}{\includegraphics[width=8pt]{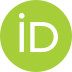}}}

\begin{document}

\preprint{APS/123-QED}

\title{Low frequency correlated charge noise measurements across multiple energy transitions in a tantalum transmon}

\author{Daniel M. Tennant \orcid{0000-0003-3688-9211}}
\email{tennant3@llnl.gov}
\affiliation{
Lawrence Livermore National Laboratory, Livermore, California 94550, USA
}

\author{Luis A. Martinez \orcid{0000-0002-2931-0833}}%
 
\affiliation{
Lawrence Livermore National Laboratory, Livermore, California 94550, USA
}%

\author{Kristin M. Beck \orcid{0000-0003-2486-4164}}
\affiliation{
Lawrence Livermore National Laboratory, Livermore, California 94550, USA
}

\author{Sean R. O'Kelley \orcid{0000-0003-0711-6471}}
\affiliation{
Lawrence Livermore National Laboratory, Livermore, California 94550, USA
}

\author{Christopher D. Wilen \orcid{0000-0002-7091-8007}}
\affiliation{
 Department of Physics, University of Wisconsin-Madison, Madison, Wisconsin 53706, USA
} 
\author{R. McDermott \orcid{0000-0001-5677-8637}} 
\affiliation{
 Department of Physics, University of Wisconsin-Madison, Madison, Wisconsin 53706, USA
}%

\author{Jonathan L DuBois \orcid{0000-0003-3154-4273}}
\affiliation{
Lawrence Livermore National Laboratory, Livermore, California 94550, USA
}

\author{Yaniv J. Rosen \orcid{0000-0003-4671-2305}}
\affiliation{
Lawrence Livermore National Laboratory, Livermore, California 94550, USA
}


\date{\today}

\begin{abstract}
Transmon qubits fabricated with tantalum metal have been shown to possess energy relaxation times greater than 300 $\muup$s and, as such, present an attractive platform for high precision, correlated noise studies across multiple higher energy transitions.  Tracking the multi-level fluctuating qudit frequencies with a precision enabled by the high coherence of the device allows us to extract the charge offset and quasi-particle dynamics.  We observe qualitatively different charge offset behavior in the tantalum device than those measured in previous low frequency charge noise studies.  In particular, we find the charge offset dynamics are dominated by rare, discrete jumps between a finite number of quasi-stationary charge configurations, a previously unobserved charge noise process in superconducting qubits.


\end{abstract}

\maketitle


\section{\label{sec:level1}Introduction}

Superconducting qubits are one of the most promising hardware platforms for near future quantum computing processors \cite{kjaergaard20}.  Recent demonstrations of high fidelity two qubit gates \cite{dicarlo09,chen14,caldwell18}, quantum simulations \cite{georgescu14,yan19,ma19}, quantum algorithms \cite{lucero12,waseem13}, and quantum error correction \cite{kelley15,ofek16,harper19,hu19} all indicate the potentiality of superconducting devices for quantum information systems.  Devices based on superconducting circuits, cavities, and resonators leverage well established fabrication techniques to create a wide variety of structures capable of the array of functionalities listed above.  While superconducting quantum devices are capable of a wide range of behavior, it is precisely this flexibility that introduces unwanted sources of noise.  Fabrication imperfections, material defects, and environmental interactions all conspire to limit the coherence properties necessary for universal quantum computation.


Much recent effort has focused on seeking material solutions to decoherence in superconducting qubits.  Energy relaxation times in these devices are believed to be limited by Two Level Systems (TLS) hosted in the superconducting metal oxide and substrate interfaces \cite{wang15,muller19,mcrae20}.  Superconducting qubits have traditionally been fabricated with either aluminum or niobium sputtered onto silicon or sapphire substrates.  Niobium, in particular, generates multiple species of oxide on its surface with significant populations extending up to 10 nm into the metal \cite{premkumar21}, creating an unwanted environment for TLS to reside \cite{cava91}.  Alternative superconducting metals might exist that produce more uniform oxides or create cleaner interfaces between the metal and substrate.

Tantalum has recently emerged as such a candidate superconducting metal \cite{place21}.  Energy relaxation times in a transmon device, fabricated with tantalum metal on a sapphire substrate, have been shown to consistently exceed 300 $\muup$s in select devices, a state of the art coherence time for the transmon design.  This indicates a reduced TLS density of states fluctuating at gigahertz frequencies in these devices.  X-ray photoelectron spectroscopy (XPS) and transmission electron microscopy (TEM) measurements of the surface reveal 2-3 nm thick layer of predominantly Ta$_2$O$_5$ which, unlike niobium, only extends 1 or 2 atomic sites into the bulk \cite{deleon21}.  Additionally, when compared to aluminum resonators and qubits with similar surface participation ratios, tantalum devices exhibit loss tangents reduced by a factor of 3-4 \cite{wang15,deleon21}.  

To further understand the coherence properties of the tantalum qubit, we perform low frequency charge noise measurements on a tantalum transmon.  Specifically, we measure both the quasi-particle tunneling rate across the Josephson junction and the low frequency charge offset dynamics of the transmon environment.  While the contribution of quasi-particle fluctuations to qubit decoherence is not believed to be a dominant factor, it does limit superconducting qubit operations on a fundamental level and thus motivates our study \cite{catelani11,catelani12,catelani14}.  The low frequency superconducting qubit charge environment is comprised of multiple processes hosted both in the device substrate and superconducting metal surface \cite{degraaf20}.  These processes are not well understood, especially in the case of tantalum at cryogenic temperatures.  

Low frequency charge noise measurements have been performed on 3D transmons \cite{riste13,serniak18,serniak19} and single island planar transmons \cite{christensen19, wilen20}.  These devices spanned a number of material and design combinations: a 3D aluminum transmon on sapphire substrate in a copper \cite{riste13} and aluminum \cite{serniak18,serniak19} cavity, and a planar, single island niobium transmon, with aluminum Josephson junctions, on a silicon substrate \cite{christensen19, wilen20}.  These previous studies all employed charge sensitive transmons such that their $f_{01}$ transition frequency was sensitive to the device's charge offset.  They all found quasi-particle dwell times of 1-6 ms \cite{note}, confirming their coherence properties are not limited by quasi-particle fluctuations.  Additionally, all previous studies observed diffusive charge environment dynamics, uniformly sampling the device offset charge, mod 2$e$.  This behavior was interrupted by large ($>$ 0.1 $e$), discrete charge jumps separated by 5-7 minutes \cite{riste13,serniak18,serniak19,christensen19} and 12 minutes \cite{wilen20} on average.  These discrete charge events have been reasonably modeled as ionizing radiation, predominantly $\gamma$-rays, colliding with the device substrate and producing liberated charges and energetic phonons \cite{christensen19,wilen20}.

For our measurements, we make use of the multi-level, or qudit, nature of the transmon.  The improved coherence of the tantalum transmon, which persists in the excited states, allows a high precision frequency measurement across multiple levels of the device.  The improved stability of the excited states enables the use of long time, narrow bandwidth excitation pulses capable of resolving the two charge-parity bands of the higher transitions.  Unlike similar noise studies previously performed \cite{riste13,serniak18,serniak19,christensen19,wilen20}, this device is fully in the transmon regime, $E_J \gg E_C$, with the charge dispersion of the 0-1 transition much less than the natural linewidth of the corresponding transition.  This is not the case for the higher transitions, and with the narrow bandwidth pulses allowed by the high coherence of the device, it is possible to resolve the two separate frequencies of the two charge-parity states down to a separation of approximately 25 kHz.  With this capability, we monitor both the quasi-particle tunneling rate across the Josephson junction as well as the charge offset dynamics affecting the qubit.  In addition, we perform these measurements in an interwoven manner across multiple transitions to detect correlated charge noise behavior.  

We find that while the quasi-particle tunneling rates observed in the tantalum transmon are similar to those measured in aluminum and niobium devices, the charge offset dynamics are qualitatively different.  The differential tantalum transmon charge offset dynamics are dominated by fluctuations between a few meta-stable configurations of the charge environment.  This leads to remarkably stable charge offsets, often lasting hours, punctuated by large discrete charge jumps.  By repeating our measurements over a range of cryogenic temperatures, it is possible to discern two categorical sources of noise.  One type of charge noise event is temperature dependent and causes predominantly nearest-neighbor charge offset transitions.  The other is temperature independent and causes large charge offsets.  The temperature independent noise compares favorably to ionizing radiation events while temperature dependent noise is likely on-chip charge fluctuations which may be a function of tantalum’s uniquely uniform surface chemistry.

\begin{figure}[t]
\centering
\includegraphics[width = \linewidth]{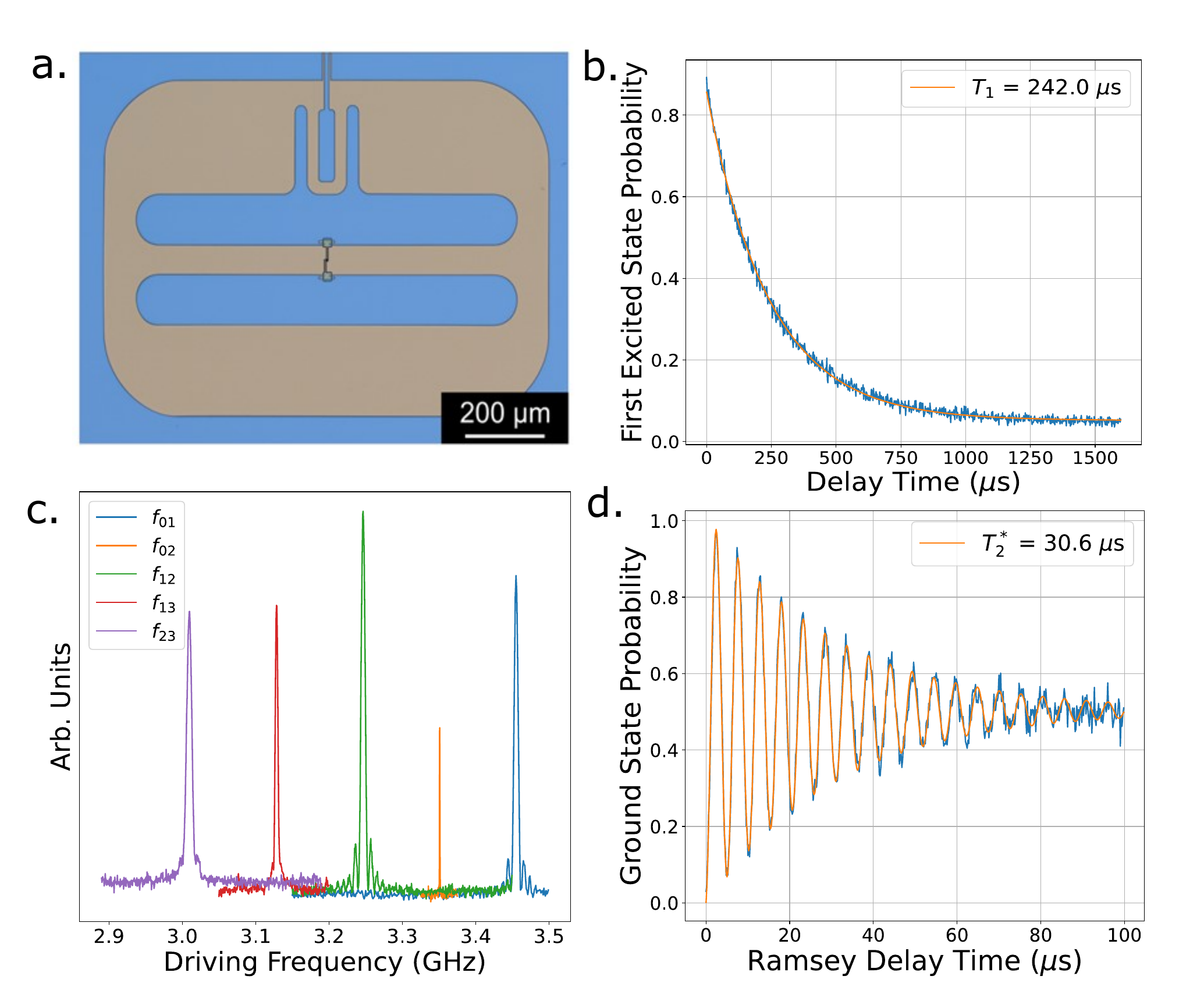}
\caption{Planar tantalum transmon: \textbf{a.}  False color optical image of differential transmon device, courtesy of \cite{place21}.  In blue are the tantalum capacitor pads, ground plane and coupling to the readout resonator.  The Josephson junction connecting the two capacitor paddles is aluminum.  \textbf{b.}  For this particular device we consistently measure energy relaxation times in the qubit subspace of 247 $\pm$ 9 $\muup$s.  \textbf{c.} Shown are spectroscopy measurements of the qudit levels we will make use of in this work.  \textbf{d.}  Ramsey measurement in the qubit subspace yield $T_2$* of approximately 30 $\muup$s, depending on cooldown.} 
\end{figure}  


\section{\label{sec:level1}Device}

The tantalum transmon used in this study was designed and fabricated by the Houck group at Princeton  \cite{place21}.  To produce the device, tantalum was sputtered on a sapphire substrate heated to 500 C to ensure growth of tantalum's $\alpha$ phase.  Photolithography and wet etching were then applied to define the capacitor and resonator structures.  Finally, electron beam lithography and an \textit{in situ} ion etch were applied followed by electron beam evaporation to create the Al/AlO$_x$/Al Josephson junction.  Between most fabrication steps, optimized solvent and piranha cleaning were used to reduce contaminants introduced during the fabrication steps.  

The Hamiltonian for the transmon in the charge basis is
\begin{equation}
\begin{split}
H = 4\,E_C\, & (\hat{n} - n_{g})^2 \ket{n}\bra{n} \\ -  
& \frac{E_J}{2} (\ket{n+1}\bra{n} + \ket{n-1}\bra{n}),
\end{split}
\end{equation}
where $\hat{n}$ is the operator representing the number of Cooper pairs on the transmon island and $n_g$ is the dimensionless gate charge in units of 2$e$, experienced by the transmon \cite{koch07}.  It is the fluctuating gate charge, $n_g$, or charge offset, that we are primarily measuring in this work.  The transmon is designed to be in the strong transmon regime, with $E_J/h = 6.3366$ GHz and $E_C/h = 208.3$ MHz, ensuring the 0-1 transition to be completely insensitive to charge offset.  In these measurements, we take advantage of the device spectrum up to the third excited state, treating the transmon as a slightly anharmonic oscillator as opposed to a two level qubit \cite{zhu13,braumuller15}.  

\begin{figure*}[t]
\centering
\includegraphics[width = \linewidth]{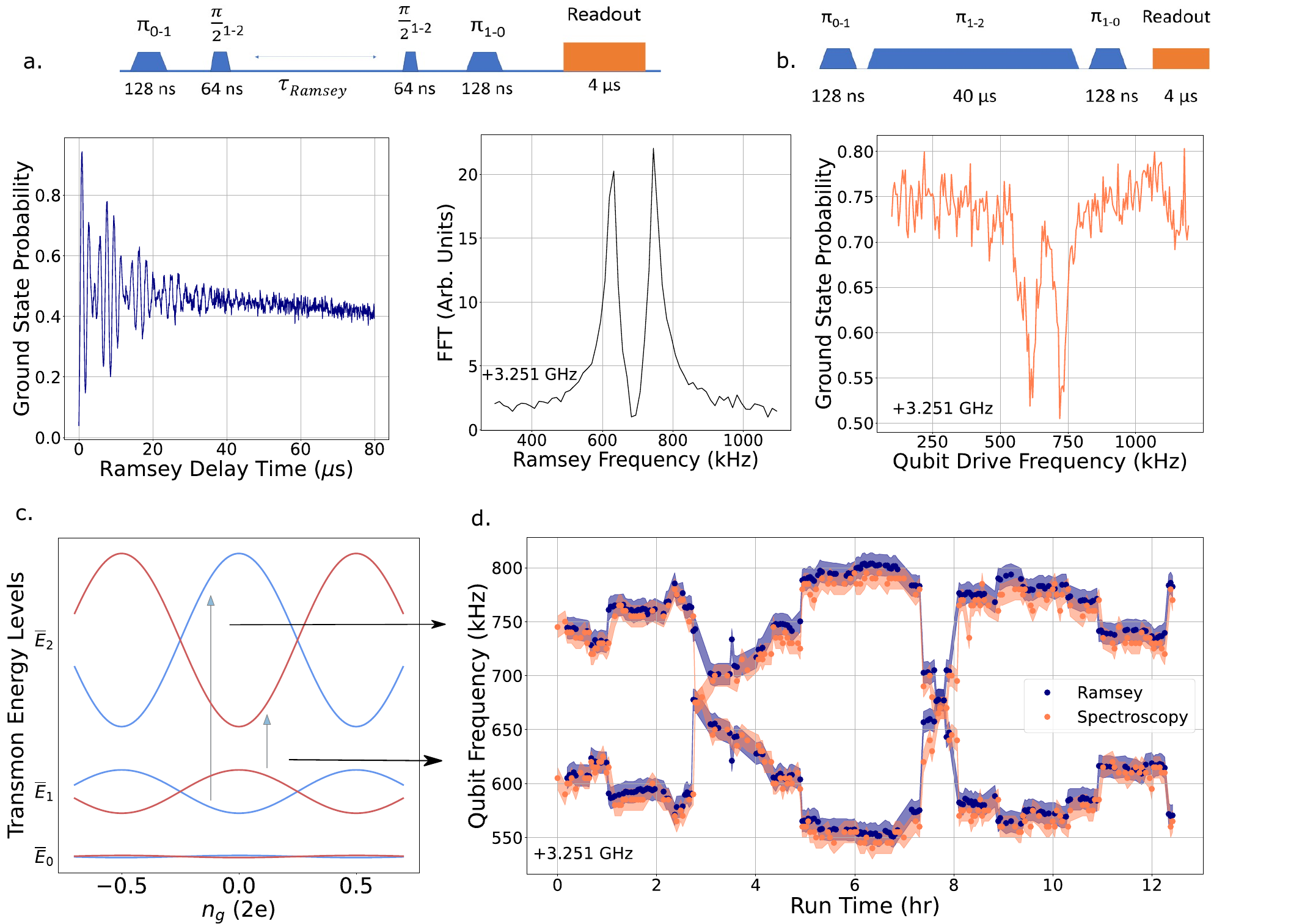}
\caption{Comparison between qubit spectroscopy and Ramsey measurements in higher levels.  \textbf{a.}  Ramsey measurements in the 1-2 qudit subspace shows a beat frequency in its decay.  A fast Fourier transform is applied to the measurement trace and the two frequencies corresponding to the two charge-parity state frequencies are revealed.  \textbf{b}.  The high precision qubit spectroscopy measurement in the 1-2 level uses a 40 $\muup$s excitation pulse after being prepared in the qudit's first excited state.  The narrow bandwidth 1-2 excitation pulse allows the two charge-parity state frequencies to be resolved.  \textbf{c.}  The energy level diagram of the qudit's excited states as a function of charge offset illustrates how the the two frequencies present in these measurements are due to the qudit being in either its even or odd charge-parity state.  \textbf{d.}  Taking interwoven qubit spectroscopy and Ramsey measurements over the course of twelve hours show clear agreement between the two methods.} 
\end{figure*}


As shown in Fig. 1, the transmon used in these measurements possesses a $f_{01}$ = 3.4578 GHz with an anharmonicity of -208.3 MHz.  Population of the higher states is achieved by both single photon sequential state population transfer between adjacent qudit states \cite{peterer15} and two photon state population transfer between next-nearest qudit states \cite{schreier08,deppe08}.  Similar to previous measurements on 3D transmons, energy relaxation predominantly occurs through a ladder of adjacent qudit states \cite{peterer15}, and due to the high coherence of the tantlaum device, even these higher levels have extended lifetimes \cite{sm}.

As previously mentioned, the charge dispersion of the 0-1 transition is too small to be detectable, but the higher transitions, starting with the 1-2 transition with a maximum charge dispersion of 120 kHz, have large enough charge dispersions as to make their two charge-parity bands distinguishable.  Quasi-particles tunnel across the Josephson junction \cite{catelani11,catelani12} causing a change in offset charge of 1$e$ at a rate much faster than the total Ramsey or spectroscopy measurement time.  As illustrated in Fig. 2c, this leads to two distinguishable frequency bands per qudit energy transition.  The separation of these two frequencies is a measure of the charge offset experienced by the qudit via
\begin{equation}
f^{\pm}_{ij} = \bar{f}_{ij} \pm \epsilon_{ij} \cos(2\pi n_g),
\end{equation}       
where $\epsilon_{ij}$ is the difference in charge dispersion between the two levels being probed, $\bar{f}_{ij}$ is the average frequency of the transition, $n_g$ is the dimensionless gate charge counted in units of 2$e$, and the $\pm$ refers to whether there is a difference of an even or odd number of quasi-particles across the Josephson junction. 

\begin{figure*}[t]
\centering
\includegraphics[width = \linewidth]{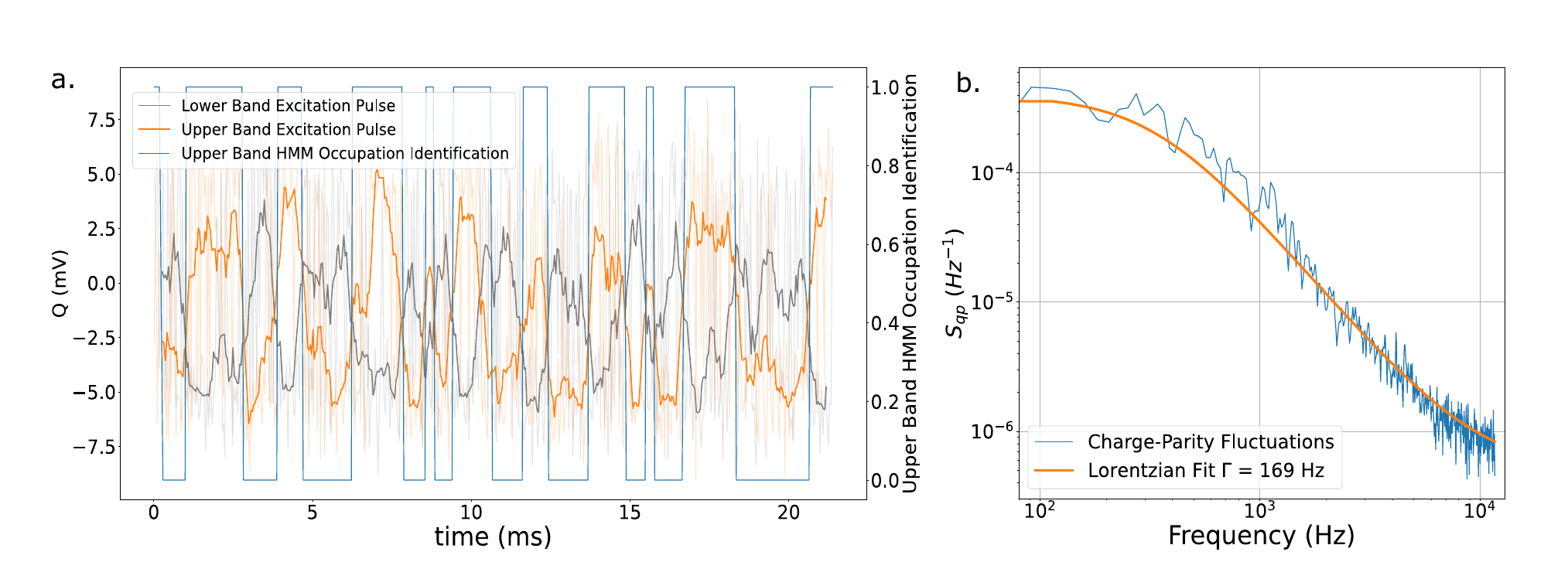}
\caption{Charge-parity fluctuations measured in the 1-2 transition.  \textbf{a.}  A mid-band qubit excitation pulse of 10 $\muup$s was applied to one charge-parity qubit state frequency followed by the opposite charge parity qubit state frequency.  These measurements followed one another with a total duty cycle of approximately 50 $\muup$s.  The left vertical side of Panel \textbf{a} indicates the quadrature channel demodulated voltages in faded color, phase tuned for high resolution between the ground and excited states, of the single shot data points.  As a guide for the eye, the 10-point running average of the output voltages are shown in bold.  The Hidden Markov model results, applied to the single shot data, are displayed on the right vertical axis.  \textbf{b.}  The quasi-particle dynamics, fit to a Lorentzian distribution on a white noise background, reveal a typical dwell time of $1/\Gamma$ = 5.9 ms. } 
\end{figure*}  


\section{\label{sec:level1}Measurements}

The exceptional lifetimes of the higher levels, combined with the exponential increase in charge dispersion with respect to qudit energy level \cite{koch07}, make the higher levels of this device a particularly well suited platform for studying low frequency charge noise.  The coherence properties of the higher levels of a 3D transmon have been previously explored using Ramsey spectroscopy \cite{peterer15}.  By applying a ladder of $\pi$ pulses, it is possible to populate the higher states with high fidelity.  Following this ladder upwards, a Ramsey measurement can be performed in a qudit subspace spanned by two adjacent levels.  As shown in Fig. 2a, by performing a fast Fourier transform on the resulting trace, the two charge-parity qudit frequencies of that subspace can then be extracted.  

In our measurements, as well as in previously reported higher level Ramsey measurements, there are typically two separate frequencies present in the Ramsey decay.  These two frequencies always fall within the expected charge dispersion of the investigated level transition.  Hence, as the total measurement time is much longer than the quasi-particle fluctuation time, this is interpreted as the experiment sampling both charge-parity states of the transmon.

This conjecture is confirmed by high precision spectroscopy measurements of the higher levels.  The lifetime of the $\ket 1$, state, for example, is approximately 247 $\muup$s.  This allows us, after populating the $\ket 1$ state with a short, high fidelity $\pi_{01}$ pulse, to apply longer $\pi_{12}$ pulses, up to the 10's of $\muup$s, without significant relaxation out of the $\ket 1$ state.  As illustrated in Fig. 2b, measurements reported in this article utilized high precision $\pi_{ij}$ pulses with a typical width of 40 $\muup$s allowing a spectral resolution of approximately 25 kHz.  Performing concurrent Ramsey and spectroscopy measurements yields the two charge-parity state frequencies in agreement within experimental uncertainties.  


Furthermore, high precision qudit spectroscopy and Ramsey measurements in the 1-2 qudit transition were taken repeatedly in an interwoven manner over the course of 12 hours.  As shown in Fig. 2d, the two sets of extracted qudit transition frequencies, corresponding to the two charge-parity states, agree well with one another over the course of many hours.  These measurements provide a confirmation to the conjecture posed in \cite{peterer15} that the two frequencies extracted by Ramsey measurements in the higher qudit levels are caused by an interplay between low frequency charge environment dynamics and millisecond scale quasi-particle tunneling.

A similar spectroscopy measurement protocol can be employed to extract the charge-parity state of the qudit.  Quasi-particle fluctuations have been measured to occur on timescales from 0.8 - 6 ms \cite{riste13,serniak19,christensen19}.  Therefore, single shot measurements and appropriate signal analysis must be employed to extract the quasi-particle dynamics.  Similar to the high precision, higher level qudit spectroscopy measurements described in the previous paragraph, we prepare the $\ket 1$ state with a short time, high fidelity $\pi_{01}$ pulse.  Then, instead of implementing a high precision, 40 $\muup$s duration pulse, we enact a moderately precise $\pi_{12}$ pulse of 10 $\muup$s.  This pulse has a spectral content of approximately 100 kHz, approximately the size of the charge dispersion of one of the charge-parity bands.  The $\pi_{12}$ qudit excitation pulse is executed with a center frequency just inside the maximum charge dispersion frequency, followed by the measurement sequence repeated with a center frequency targeted at the opposite charge-parity band.  Each measurement sequence is punctuated by an active qubit reset protocol, enabled by the digitizer's on-board field-programmable gate array (FPGA), allowing a total duty cycle of approximately 50 $\muup$s \cite{sm}.  These measurements are then repeated, in an interwoven fashion, for a minute.

\section{\label{sec:level1}Results}

Shown in Fig. 3 are the results of the quasi-particle tunneling measurement.  A 25 ms portion of the entire minute long trace is displayed in Panel a.  In faded color is the raw single shot demodulated quadrature voltage data, phased adjusted to be most sensitive to the transition.  A 3-state general mixture model is then applied to the single shot data to convert the demodulated I/Q voltages to qubit state classification probabilities.  Finally, a Hidden Markov model (HMM) was then applied to the state classified traces to produce the charge-parity transitions show on the right vertical axis of Fig. 3a \cite{sm}.  As a guide to the eye as well as further validating our HMM analysis, the ten-point running average of the quadrature voltage is also displayed.    

The Power Spectral Density (PSD) is taken of the charge-parity state occupation trace.  This PSD is well fit by a Lorentzian curve with a characteristic frequency of 169 Hz.  This corresponds to a typical quasi-particle dwell time of $1/\Gamma$ = 5.9 ms, consistent with other reported results on devices with aluminum Josephson junctions \cite{riste13,serniak19,christensen19}.  As noted by previous authors, we take this observation of quasi-particle dwell time being greater than the energy relaxation time of the qudit to be evidence that quasi-particle tunneling is not the dominant mechanism limiting relaxation times in the tantalum device \cite{catelani14,serniak18,serniak19}.      

In order to characterize the transmon's charge environment dynamics, interwoven qudit spectroscopy measurements were performed across the 1-2 single-photon and 1-3 two-photon qudit transitions repeatedly over the course of approximately 70 hours.  Both transitions have charge dispersions that agree well with simulations of the device.  Taking into account the full charge dispersions of the two transitions, and the relationship in Eq. 2, both frequency traces can be converted to a charge offset.  Due to the lack of charge offset control in this device, we can only project the measured frequencies onto the 0-0.5 $e$ range.  A strong agreement in the charge offsets extracted from the 1-2 and 1-3 transitions, illustrated in Fig. 4a, confirms that both transitions are influenced by a common charge environment.

Looking closer at the charge offset distribution of this long time trace reveals some unexpected features.  Previous measurements of the charge offset distribution yield near uniform charge distributions \cite{riste13}.  The charge environments sampled all possible offset values evenly, not preferring one charge offset over others.  The local charge environment of the tantalum transmon behaves quite differently.  As shown in Fig. 4b, there are charge offsets the system prefers and regions of charge offset the system only very rarely samples.  The charge offset dynamics are dominated by long periods, up to hours, of stability.  These long stable periods are punctuated by large discrete charge jumps between these quasi-stable charge configurations.  The fact that the frequency traces across multiple transitions, when converted to a common charge offset, behave quantitatively similarly, indicate that this non-uniformity is not due to a resonant or dispersive frequency interaction, but rather is a characteristic of the transmon's charge environment.

To further investigate the low frequency charge dynamics, we collected the charge offset values into the separate time bins shown in Fig. 4c.  The different colors represent different 17 hour consecutive sets of data.  If the system is somehow not in equilibrium and we allowed it to evolve for a long enough time, the charge offset distribution would approach uniformity.  If this were so, the different colored data sets in Fig. 4c would be quantitatively distinct from each other as the charge environment further explored its configuration space.  This is not the case.  The different colored data sets appear fairly self-similar with no new favored locations appearing at later times.  This leads us to believe the charge environment visible to the tantalum transmon has fully sampled its configuration space in the first 17 hours. 

\begin{figure*}[t]
\centering
\includegraphics[width = \linewidth]{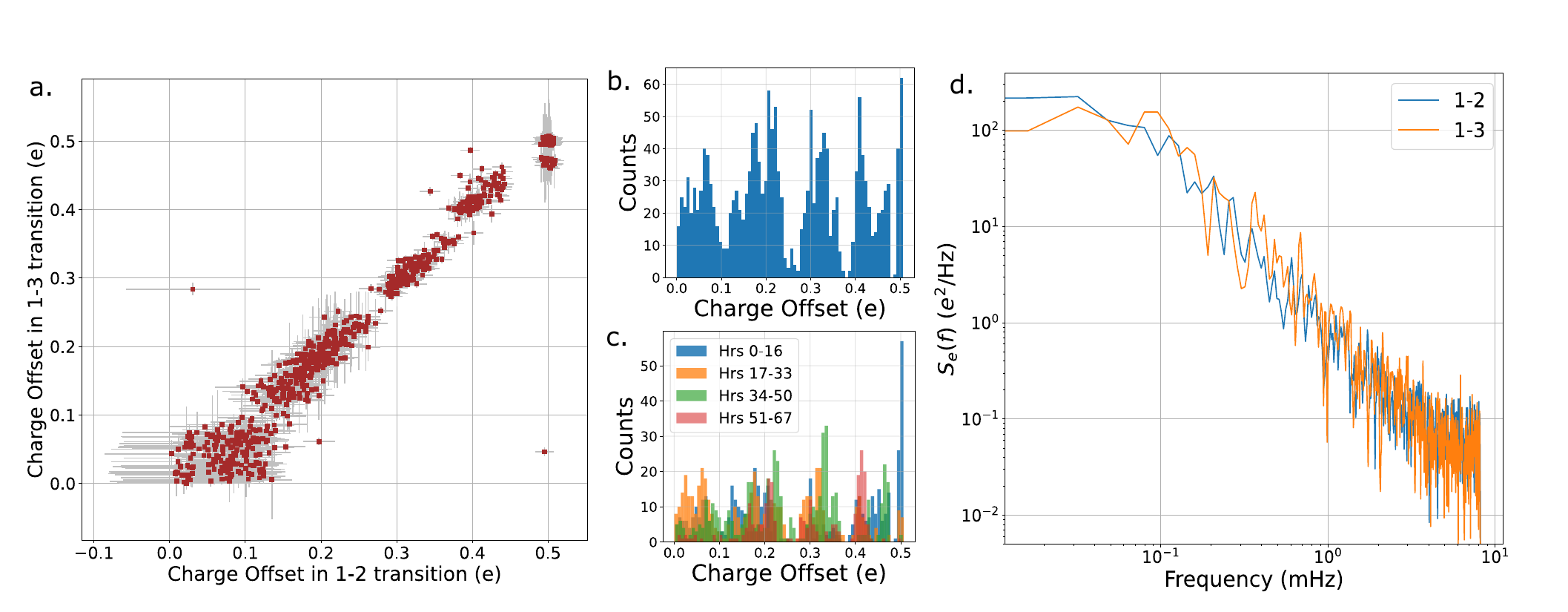}
\caption{Low frequency dynamics of the qubit charge environment.  \textbf{a.}  High precision qubit spectroscopy measurements in the 1-2 and 1-3 transitions were taken interwoven with each other over the course of 70 hours.  Displayed are ordered pairs of the offset charge extracted from adjacent measurements of the 1-2 and 1-3 transitions showing a clear correlation between the two data sets.  \textbf{b.}  The  charge offsets form a highly nonuniform distribution.  \textbf{c.}  The same data displayed above in Panel \textbf{b} is binned such that the different colors in the histogram correspond to each 17 hours of data.  The similarity between the different time bins suggests the charge environment prefers certain discrete configurations as opposed to simply not having the opportunity to fully explore its configuration space in the measurements time.  \textbf{d.}  The power spectral density of the charge offset dynamics at base temperature shows an $1/f^\alpha$ noise with $\alpha$=1.94 and magnitude at 1 Hz of $1.11 \times 10^{-6}\,e^2/\rm{Hz}$.} 
\end{figure*} 


Displayed in Fig. 4d is the PSD of the low frequency charge dynamics at base temperature.  It yields a $1/f^\alpha$ spectrum with $\alpha_{\rm{e}}$ = 1.94 and magnitude at 1 Hz of $1.11 \times 10^{-6} \, e^2$/Hz.  Choosing instead to view this as frequency noise, by analyzing the PSD of the original frequency trace, allows us to make a direct comparison to previous charge noise measurements.  By normalizing our charge dispersion to a common frequency with previous measurements, we find a similar character of noise spectrum with $\alpha_{\rm{f}} = 2.06$.  However, our measurements yield a charge noise magnitude at 1 Hz two orders of magnitude smaller, $A = 7.4 \times 10^5$ Hz$^2$/Hz, than previous measurements on aluminum and niobium devices, which consistently fall in the $10^7$ Hz$^2$/Hz range \cite{riste13,serniak18,christensen19}.  

We then repeated the long time series of measurements of 1-3 two photon qudit spectroscopy measurements at 50 mK, 100 mK, and 150 mK.  For each temperature, we employed a HMM to identify the finite number of quasi-stable states and their occupation time.  We repeated this analysis forcing the stable charge configurations to be fixed across temperatures and, alternatively, allowing them to be independently determined at each temperature.  The two methods yield similar results for the number and location of quasi-stable charge offsets \cite{sm}.  For the remainder of the paper we will use the charge offsets identified by analyzing the data sets from all temperatures simultaneously.  The states displayed in Fig. 5a-d are labeled sequentially 1-8 ranging from charge offset 0-0.5 $e$.  Shown in Fig. 5e are the average stable charge configuration times for each temperature.  This shows a clear decrease of stable charge time with increasing temperature.  

\begin{figure*}[t]
\centering
\includegraphics[width = \linewidth]{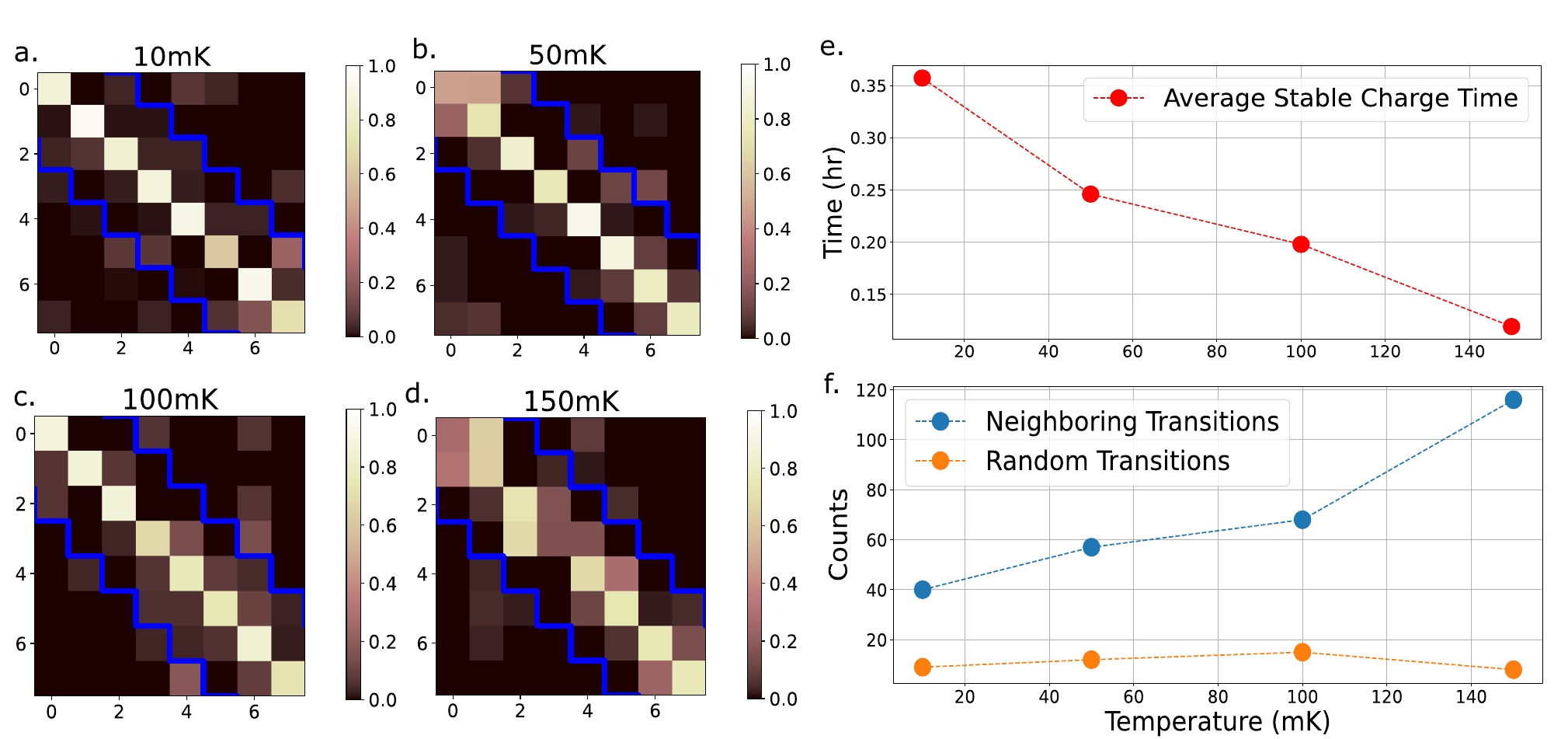}
\caption{The transition matrices between quasi-stable charge configurations at various temperatures.  A HMM was employed to identify the finite number of quasi-stable charge configurations consistently across temperatures.  \textbf{a-d.}  The diagonal elements of the matrices represent the probability for the charge environment to remain in that same configuration.  The off-diagonal elements represent the probability for the charge environment to transition between configurations.  The region denoting nearest and next-nearest neighbor transitions are outlined in blue.  For each temperature, the majority of transitions happen between adjacent charge configurations.  \textbf{e.}  Shown are the average stable charge times for each temperature.  They show a decrease in stable charge time with increasing temperature.  \textbf{f.}  The transitions can be neatly divided into nearest and next-nearest neighbor hopping and the remaining transitions.  The nearest and next-nearest hopping shows a increasing dependence on temperature while the remaining transitions are temperature independent.} 
\end{figure*}  


The transitions between states, or off-diagonal elements in Fig. 5a-d, can divided neatly into temperature dependent and independent subgroups.  As shown in Fig. 5e, the nearest and next-nearest state transitions increase with temperature while the remaining transitions are temperature independent.  This suggests separate physical mechanisms for the two subgroups of transitions, i.e. thermally assisted tunneling between similar charge configurations being responsible for the neighboring transitions and some external noise source, independent of mixing chamber temperature, being responsible for the charge configuration scrambling transitions.

To summarize our results, we have measured an approximate $1/f^2$ low frequency charge noise PSD with amplitude two orders of magnitude smaller than previous measurements on aluminum and niobium superconducting qubits.  The previous studies were based on devices fabricated in both planar and 3-D geometries as well as having both silicon and sapphire substrates.  Also distinct in our results is that the low frequency charge noise study on the tantalum qubit revealed preferred charge offset values indicating a finite number of quasi-stable environmental charge configurations.  This is different from all previously reported low frequency charge noise measurements performed on superconducting qubits regardless of their geometry or substrate.


Large discrete charge offset jumps have been recently reported in niobium devices with aluminum Josepshon junctions fabricated on a silicon substrate with typical occurrence times of approximately 12 minutes \cite{wilen20}.  The discrete charge jumps measured in these niobium transmons were reasonably modeled as gamma ray collisions with the device substrate.  The average time between charge offset jumps in the tantalum device at 10 mK is almost twice as long, approximately 22 minutes, with some stable charge times reaching over 4 hours.      

To further investigate whether our data can be explained in terms of these ionizing particle tracks, we performed finite element electromagnetic simulations of both the niobium transmons used in this previous study and the tantalum transmon to determine the volume of the substrate that the transmons are sensitive to charge events such as gamma ray collisions \cite{sm}.  While the differential design of the tantalum does reduce its charge sensitive region into its substrate, the overall much larger size of the tantalum transmon compared to the single island niobium device actually makes the tantalum transmon's charge sensitive region larger.  This would suggest, all other things being equal, that the tantalum device's charge environment would fluctuate at a faster rate. 

This discrepancy becomes even more acute if we compare the previously reported rate of charge events occurring every 0.2 hours with our temperature independent charge rate of an event every 1.6 hours.  Some of this difference can be explained by the greater bandgap of sapphire, 8.3 eV, compared to that of silicon, 1.2 eV.  However, if the scrambling charge transitions observed in the tantalum transmon's charge environment are due to particle collisions with the sapphire substrate, the tantalum-sapphire device has a much reduced characteristic radiation trapping and charge production rate.

The temperature dependent behavior of the neighboring charge configuration transitions point to these transitions being caused by jumps between local charge configurations in the immediate vicinity of the transmon.  The fact that the temperature dependent transitions are between neighboring configurations also seems to rule out a collection of Two Level Systems (TLS) as the basis of these quasi-stable charge offsets.  TLS have been identified as a primary source of charge noise in Single Electron Transistors (SET), devices typically fabricated out of aluminum on silicon substrates \cite{farmer87,zorin96,zimmerman97,kenyon00}.  While the amplitude of charge noise found in these studies is of the same order of magnitude measured in our device, there are important qualitative differences.  The low frequency charge noise typically measured in SETs is $1/f$ in character, points to a collection of TLS with a broad range of frequencies as the underlying mechanism.  Additionally, if TLSs were primarily responsible for the temperature dependent transitions measured in our study, the transitions would be distributed across the matrices shown in Fig. 5a-d as opposed to being concentrated around the diagonal.  For example, if the eight quasi-stable charge offsets were due to three ($2^3=8$) independently fluctuating TLS, there would be three populated transitions per row in Fig. 5a-d evenly distributed across the transition matrices.  To further test this hypothesis, it could be possible to design superconducting devices capable of spatially detecting local charge fluctuations.  In fact, local charge dynamics have recently been tracked in the vicinity of a multi-mode superconducting qubit \cite{wills21}.

One possible explanation of the qualitatively different behavior of the tantalum transmon's charge environment is the different surface chemistries found in the superconducting device metals.  Tantalum has been shown to produce a highly uniform Ta$_2$O$_5$ oxide layer which could greatly reduce the density of surface defects and suppress charge tunneling between these localized states.  High quality interfaces have also been demonstrated in the tantalum device \cite{place21} which could play a role in the reduction of surface defects.  Niobium and aluminum host a much more irregular oxide layer, possibly allowing a high density of closely packed defect sites with low tunneling barriers to facilitate charge transport across its surface. 

All of this discussion points to the need of systematic comparisons between superconducting qubits of different designs, materials, and substrates as well as first principle calculations to better understand and mitigate charge noise processes in these devices.  For example, there is a growing appreciation of the detrimental effects of radiation on qubit coherence properties \cite{vepsalainen20,wilen20,cardani21,mcewen21,karatsu19}.  If the sapphire substrate does, in fact, effectively reduce the charge production rate due to ionizing particle collisions, it might be possible to accommodate the stringent requirements of surface error correction codes \cite{martinis21}.

\section{\label{sec:level2}Conclusions}

Taking advantage of the long coherence times of a tantalum based superconducting qudit, we have performed high precision frequency measurements of the higher levels of this device.  Interweaving both Ramsey and spectroscopy measurements across multiple transitions, we have explicitly confirmed previous conjectures on the charge noise behavior in the transmon's higher levels.  In particular, by performing interwoven Ramsey and spectroscopy measurements in the higher qudit levels over the course of many hours, we have confirmed that the fluctuating Ramsey beat frequency found in this and previous measurements \cite{peterer15} is due to a sampling of the two charge-parity state frequencies.  Also, by tracking the two fluctuating frequencies in multiple transitions confirms the assumption that multiple transitions do experience a common charge environment.        

Most striking of our findings however is the novel behavior of the tantalum transmon's charge environment.  There appear to be a finite number of quasi-stable configurations the charge environment samples.  This is at odds with all previous low frequency charge noise studies on superconducting qubits.  Additionally, the transitions between charge configurations happen at a much lower rate than discrete jumps in charge offsets observed in previous measurements on aluminum and niobium transmons.  If our measured discrete charge offset jumps are in fact due to ionizing particles colliding with the sample substrate, this implies that the tantalum qubit-sapphire substrate device experiences either a reduced charge production rate or a reduced released charge trap length.  However, the temperature dependence of the stable charge configurations, and their transition behavior, further complicates this picture.  The increased rate of transitions between charge configurations as a function of temperature, as well as the fact that they happen primarily between adjacent configurations, suggest that the local chemistry of the device is responsible for the majority of the behavior we observe.     

The long coherence times of the superconducting tantalum offer exciting technological possibilities for utilizing higher qudit levels.  With coherence times in the hundreds of microseconds, tantalum based qudit devices could offer the first platforms to realize recently proposed computational tasks that take advantage of the qudit connectivity such as more compact error correction codes \cite{campbell14}, efficient cryptology protocols \cite{bechmann00,bruss02}, and many body quantum chaos simulations \cite{smith19}.  In addition, the use of highly coherent excited states offer the possibility of performing \textit{in situ} studies of the charge noise environment in a quantum processor constructed out of these devices and eliminates the need for specialty charge-sensitive transmons, specifically designed for charge noise studies.  In closing, we believe that performing high precision frequency measurements across multiple qudit transitions can be a unique and valuable tool for distinguishing between different material sources of decoherence in superconducting devices.       

\section{\label{sec:level2}Acknowledgments}

The authors wish to thank Andrew Houck for providing the tantalum transmon used in this study and Kevin Osborn for helpful discussions on early charge noise measurements.  We thank MIT-Lincoln Laboratories and IARPA for providing the Traveling Wave Parametric Amplifier used in this study.  This work was supported by the U.S. Department of Energy, Office of Science, Office of Basic Energy Sciences under Award DE-SC0020313.  Additionally, this work was performed under the auspices of the U. S. Department of Energy by Lawrence Livermore National Laboratory under Contract No. DE-AC52-07NA27344. Experimental characterization was made possible by the National Nuclear Security Administration Advanced Simulation and Computing Beyond Moore’s Law (NA-ASC-127R-16) program support of the LLNL Quantum Design and Integration Testbed (QuDIT).


\bibliography{Ta_lowfreqnoise}

\begin{thebibliography}{51}
\bibitem [{not()}]{note}%
  \BibitemOpen
  \href@noop {} {}\bibinfo {note} {This is true except for \cite{serniak18}
  where both the quasiparticle dwell time and energy relaxation time were
  approximately 80 $\muup$s. This was attributed to a lack of high frequency
  filtering in the experimental setup.}\BibitemShut {Stop}%
\end{thebibliography}

\end{document}


\preprint{APS/123-QED}

\title{Supplementary Material for Low frequency correlated charge noise measurements across multiple energy transitions in a tantalum transmon}

\author{Daniel M. Tennant}
\email{tennant3@llnl.gov}
\affiliation{
Lawrence Livermore National Laboratory, Livermore, California 94550, USA
}

\author{Luis A. Martinez}%
 
\affiliation{
Lawrence Livermore National Laboratory, Livermore, California 94550, USA
}%

\author{Kristin M. Beck}
\affiliation{
Lawrence Livermore National Laboratory, Livermore, California 94550, USA
}

\author{Sean R. O'Kelley}
\affiliation{
Lawrence Livermore National Laboratory, Livermore, California 94550, USA
}


\author{Christopher D. Wilen}
\author{R. McDermott} 
\affiliation{
 Department of Physics, University of Wisconsin-Madison, Madison, Wisconsin 53706, USA
}%

\author{Jonathan L DuBois}
\affiliation{
Lawrence Livermore National Laboratory, Livermore, California 94550, USA
}

\author{Yaniv J. Rosen}
\affiliation{
Lawrence Livermore National Laboratory, Livermore, California 94550, USA
}

\maketitle

\beginsupplement

\tableofcontents

\newpage

\section{\label{sec:level1}Experimental Setup}

Our experimental setup, shown in Fig. S1, employs a heterodyne single-sideband modulation of a local oscillator for the readout and qudit control pulses.  We use two separate control channels to cleanly span the 0-1, 1-2 and 2-3 qudit transitions, which span about 420 MHz.  Each control channel comprises a pair of arbitrary waveform generators that are combined with a local oscillator to reach the qudit transition frequencies.  Both the I/Q arbitrary waveform generators and the analog-to-digital converter are housed within a Quantum Machines Operator-X control module.  This enables the fast feedback, active reset protocols used in the quasi-particle tunneling rate measurement.  As shown in Fig. S2, the demodulated readout tones fall into well defined groups depending on the qudit state.  The onboard FPGA of the QM OPX ADC can trigger qudit control pulses based on what quadrant the demodulated I/Q value the readout tone returns.  A series of conditional pi pulses in the 1-2 and 0-1 transitions applied after each measurement sequence ensures with high probability that the qudit begins the next measurement cycle in its ground state.

\section{\label{sec:level1}Coherence Measurements in the Higher Qudit Levels}

Before performing the high precision qubit spectroscopy pulses utilized in the low frequency charge noise measurements, the coherence properties of the higher levels of the tantalum transmon must be characterized.  Shown in Fig. S3 are the energy relaxation and Ramsey measurements taken up to the third excited state.  As shown, the exceptional energy relaxation times extend to the higher levels enabling the use of 40 $\mu$s long, narrow bandwidth qubit excitation pulses.  Ramsey measurements were performed by preparing coherent superpositions of the higher states.  Dephasing times extracted from these measurements in the 0-1 manifold, $T_1 = 247 \pm 9 \,\muup$s and $T_2^* = 31 \pm 3 \,\muup$s, fall far short of being $T_1$ limited.

\section{\label{sec:level1}Hidden Markov Model Analysis of Quasi-Particle Tunneling Measurement} 

To circumvent spurious parity flip transitions observed in the state-classified single-shot readout data, hidden Markov models \cite{schreiber17} were used to identify the charge-parity fluctuations as a function of time (see Fig. 3a in text).  First, a 3-state general mixture model (GMM) classifier trained on the IQ distributions classified the individual single-shot readout measurements for both parity bands.  The GMM classification process converted the demodulated IQ array into a state-classified array comprised of integer values corresponding to the qubit readout state for each of the parity band measurements.  A simple identification routine then labeled the case when the qubit was in the $\ket{1}$ or $\ket{2}$ state with ``1'', suggesting the qubit was in the corresponding parity band, and ``0'' if the qubit was in the $\ket{0}$ state.  This seemingly overcounting of the ``1'' state is due to possible state relaxation out of the qudit $\ket{2}$ state during measurement.  Additionally, state classification of noisy single-shot measurements exhibited spurious parity flip transitions and without additional processing the quasi-particle tunneling rate would be overestimated.  Therefore, 2-state hidden Markov models using discrete distributions were trained via unsupervised learning on each of the filtered parity-band datasets.  Next, the corresponding hidden Markov models were used to predict the most probable path for each dataset.  The predicted paths were plotted against the corresponding timestamps and the time between transitions was extracted by identifying the transitions between parity bands. 

\section{\label{sec:level1}Hidden Markov Model Identification of Discrete Charge Configurations}

Hidden Markov models (HMM) were used to identify the observed discrete charge configurations. The analysis procedure was as follows. First we combined the data for all temperatures into a single complete dataset to be used for learning the distributions of the discrete charge states. To determine the optimal number of charge states to be used in the HMM we iterated the number of distributions in a general mixture model (GMM) from $N = 1$ to $N = 19$, a sampling of which are shown in Fig. S4.  We then calculated three fitting metrics for each of the iterations; the silhouette score, distance between train and test GMMs, and the gradient of the Bayesian information criterion (BIC). This analysis helped identify the optimal number of discrete charge states while avoiding over-fitting the data. As shown in Fig. S5, we concluded $N = 8$ was the optimal number of discrete charge states.  This is shown in the silhouette score by $N=8$ being the last point before the scores level off.  For the distance between train and test GMMs and the BIC score, the points are also fairly constant following $N=8$.  Once we determined the optimal number of charge states to use for our HMM, the 8-state HMM was used to determine the transition matrices at all four temperatures.  The results of this analysis are shown in Fig. S6. Generally, as the temperature increases, the transitions between different charge states increase.  Additionally, as the temperature increases, the percentage of nearest and next to nearest neighbor transitions also increase, leaving an approximately constant number of other transitions. 

\begin{figure}
\centering
\includegraphics[width = 0.9\linewidth]{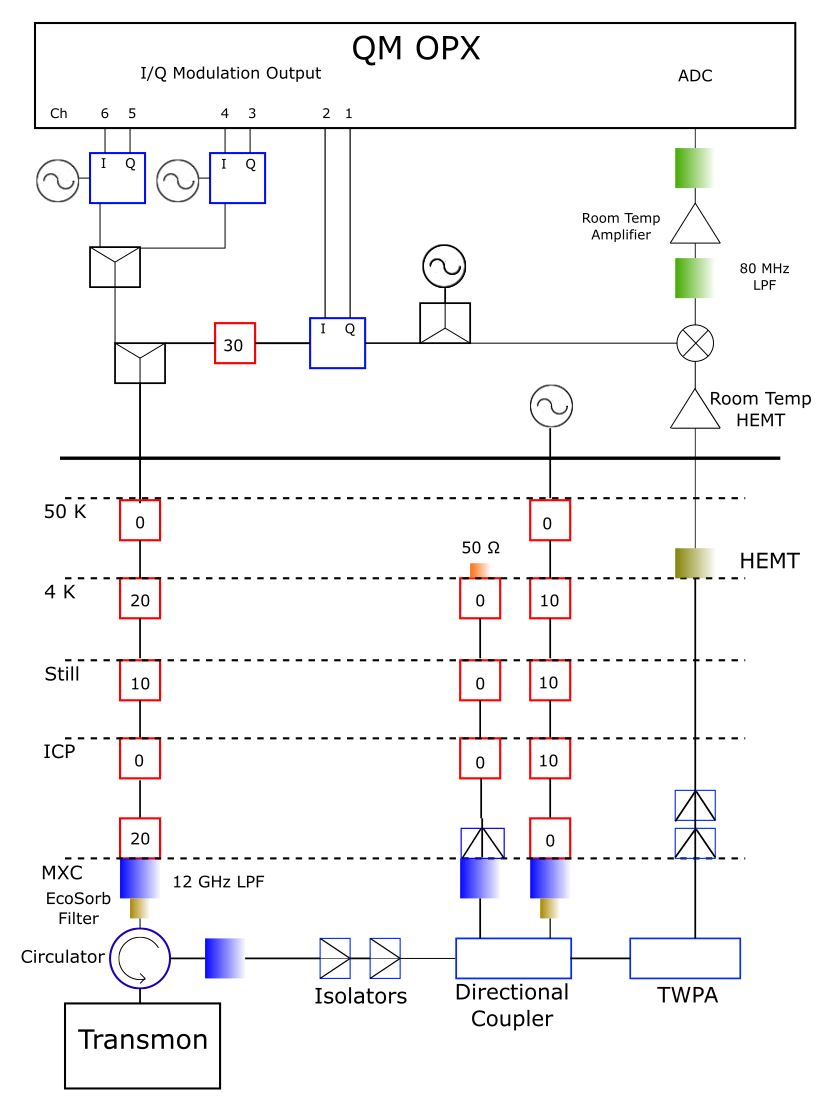}
\caption{Wiring Diagram of the Experiment.  The red boxes represent attenuators with amount of attenuation in dB indicated.  The ecosorb, in faded yellow, filters out radiation above 20 GHz.  Low pass filters of 12 GHz cutoff are in faded blue and 80 MHz cutoff are in faded green.  The transmon is enclosed in a gold-plated copper can lined with mu metal.  Microwave isolators follow both the transmon and traveling wave parametric amplifier to prevent back contamination on the Josephson elements.  The four local oscillator signals are provided by a single Holworth HSX signal generator.  All components are phase synced with a 10 MHz rubidium clock.} 
\end{figure}

\begin{figure}
\centering
\includegraphics[width = \linewidth]{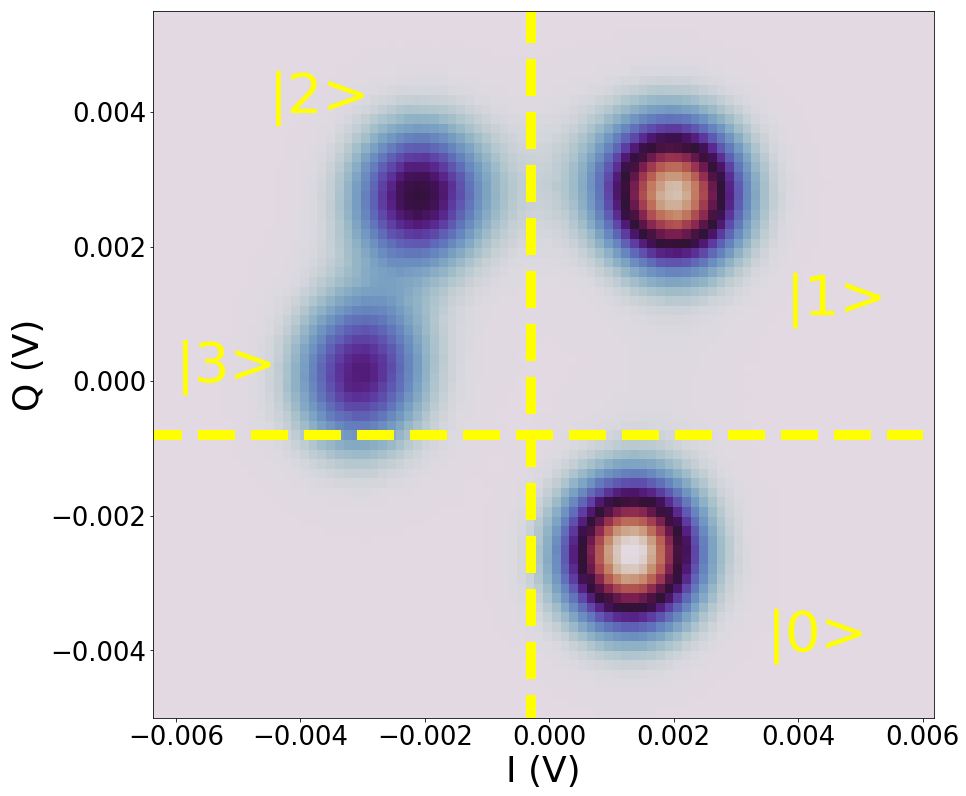}
\caption{Two Dimensional Histogram of Demodulated Readout Tone.  The four clusters in the I/Q plane correspond to the readout resonator response when the transmon is in the identified state.  The dashed lines represent conditional boundaries the digitizer FPGA uses to decide whether to send active reset pulses to the transmon between measurements.} 
\end{figure} 


\begin{figure}
\centering
\includegraphics[width = \linewidth]{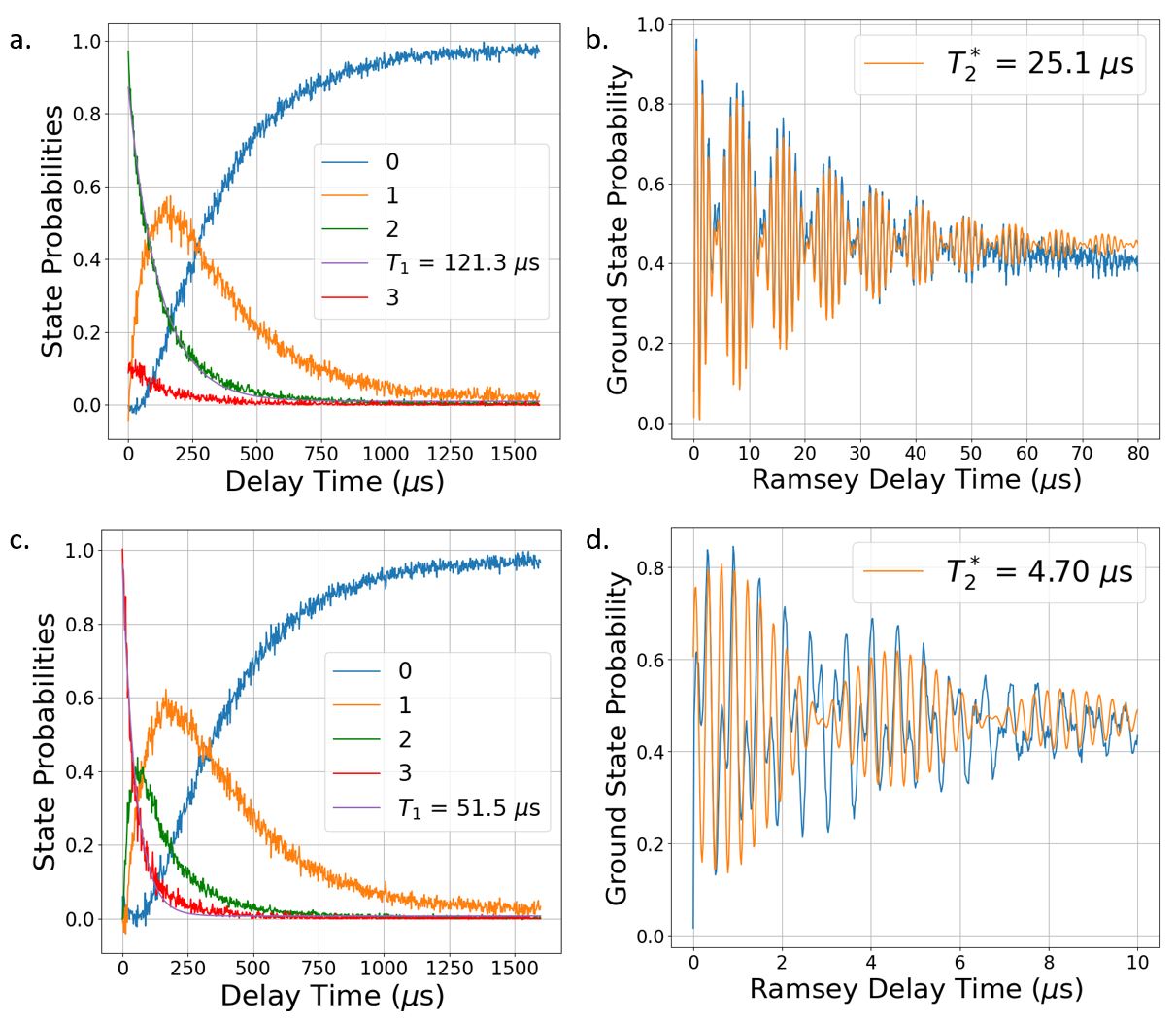}
\caption{Energy relaxation and Ramsey measurements in the second and third excited states of the transmon.  For the energy relaxation measurements, a series of $\pi$ pulses, $\pi_{01}$, $\pi_{12}$, and $\pi_{23}$ for panel \textbf{c} populate the target level.  The pulse sequence for the Ramsey measurements is $\pi_{01}$, $\frac{\pi_{12}}{2}$, Ramsey delay time, $\frac{\pi_{12}}{2}$, $\pi_{01}$ for panel \textbf{b} and $\pi_{01}$, $\pi_{12}$, $\frac{\pi_{23}}{2}$, Ramsey delay time, $\frac{\pi_{12}}{2}$, $\pi_{12}$, $\pi_{01}$ for panel \textbf{d}.  Shown in \textbf{a \& b} is the energy relaxation out of the second excited state and a Ramsey measurement demonstrating the coherence of the $\frac{1}{\sqrt{2}}(\ket{1} + \ket{2})$ state.  Shown in \textbf{c \& d} is the energy relaxation out of the third excited state and a Ramsey measurement demonstrating the coherence of the $\frac{1}{\sqrt{2}}(\ket{2} + \ket{3})$ state.  Note the sequential nature of the energy relaxation from the higher levels.  Also, the higher level Ramsey measurements typically have two frequencies present corresponding to the transition's two charge-parity bands.} 
\end{figure}

\begin{figure}
\centering
\includegraphics[width = \linewidth]{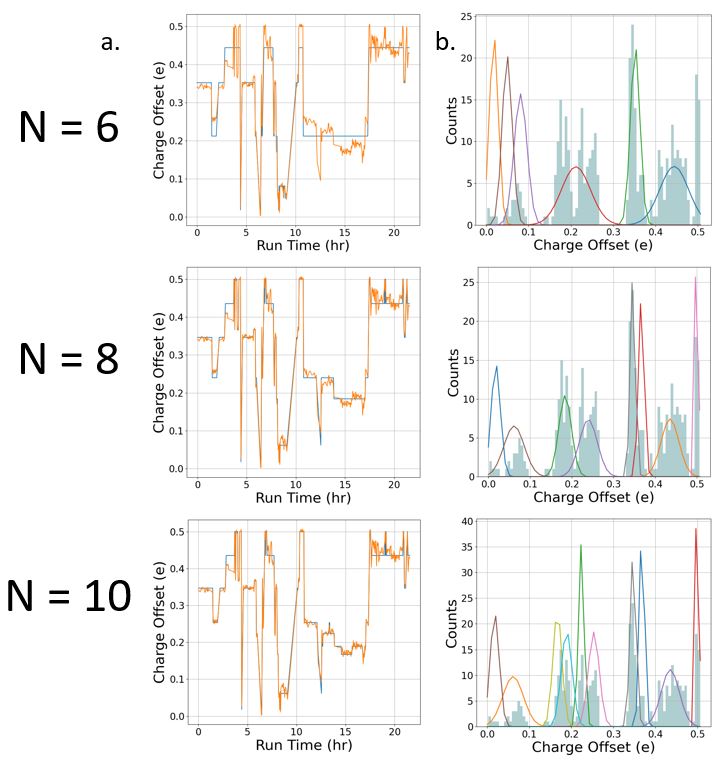}
\caption{Gaussian Mixture Model identifying the stable charge configurations with variable cluster number: \textbf{a.}  Shown in the left column are the time traces of the offset charge with different numbers of clusters as input.  Raw data is shown in yellow and the HMM identified charge configurations are in blue.  \textbf{b.} Shown in the right column are the corresponding histograms of the total charge offset results with the corresponding number of clusters identified.} 
\end{figure} 

\begin{figure}
\centering
\includegraphics[width = 0.5\linewidth]{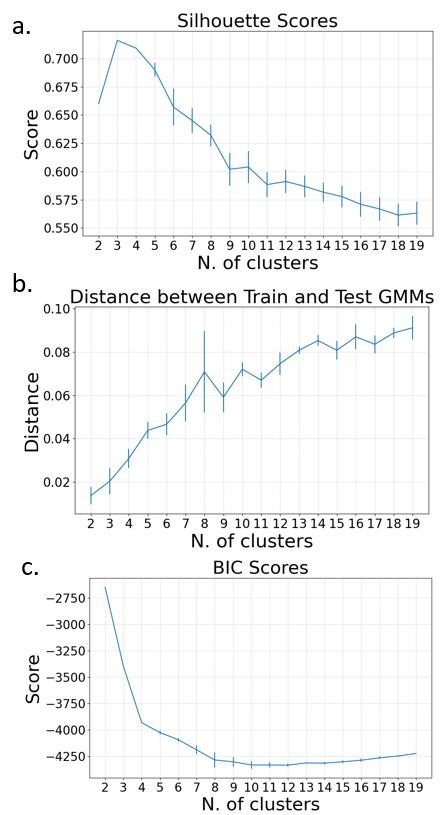}
\caption{Cluster identification analysis: \textbf{a.}  Shown is the Silhouette score for the stable charge offset identification as a function of cluster number.  Scores are relatively high until  after $n=8$.  \textbf{b.}  Shown is the cumulative distance between the data and identified charge states.  The rate of score increase slows after $n=8$.  \textbf{c.}  Shown is the Baysian score as function of cluster number.  The score plateus at $n=8$.  These different tests show $n=8$ is the most probable number of stable charge states.} 
\end{figure}   

\begin{figure*}[h]
\includegraphics[width = .9\textwidth]{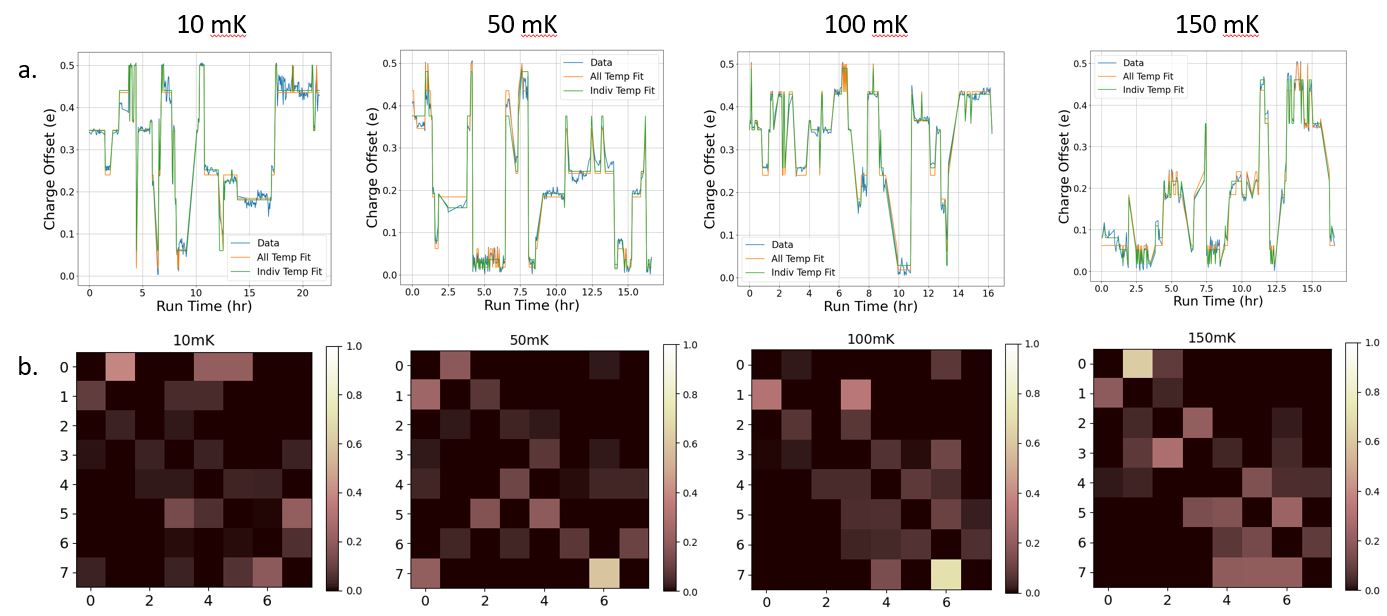} 
\caption{Temperature dependence of the low frequency charge offset dynamics.  \textbf{a.}  The charge offset experienced by the qubit is displayed along the top row for four different temperatures, 10, 50, 100, and 150 mK along with the $n=8$ cluster identification.  The $N=8$ cluster identification is shown for fitting both the individual temperature data and the total data set including all temperatures.  From the strong agreement between both fits, we conclude that the charge clusters are temperature independent and all further analysis will use the cluster identification found by fitting across all temperatures.  \textbf{b.}  Along the bottom row, the corresponding transition matrices are displayed.  The diagonal elements, which would represent the probability the charge offset remains in that particular cluster, are set to zero for better contrast in the remaining elements.  The off-diagonal elements represent the probability the charge offset transitions from one stable configuration to another.  Note that the relative probability for transitioning between adjacent states grows with temperature.}
\label{fig:HMM_charge_states}
\end{figure*}

\section{\label{sec:level1}COMSOL Device Simulations}

Recent low frequency charge noise measurements \cite{wilen20} point to gamma ray collisions with the substrate as a reasonable cause of discreet jumps of the charge offset experienced by their qubit.  As the jumps in their transmon's charge offset appear very similar to ours, albeit with a higher frequency, we explored this possible connection further.  As illustrated in Fig. S7, we performed finite element electromagnetic simulations of both the transmon in \cite{wilen20} and our own to determine the charge sensitive volume surrounding our devices.  More precisely, by placing a charge at each lattice point in the substrate, as if it had been liberated by a gamma ray collision,  what charge would it induce on the transmon island.  We found that the charge sensitive volume of the tantalum transmon was, in fact, larger than the volume experienced by the niobium device used in \cite{wilen20}.  So, if all other things were equal, and these charge offset jumps in our measurements were caused by gamma ray collisions, the tantalum device should have experienced a higher frequency of jumps, which we do not observe.

\begin{figure*}[h]
\includegraphics[width = .9\textwidth]{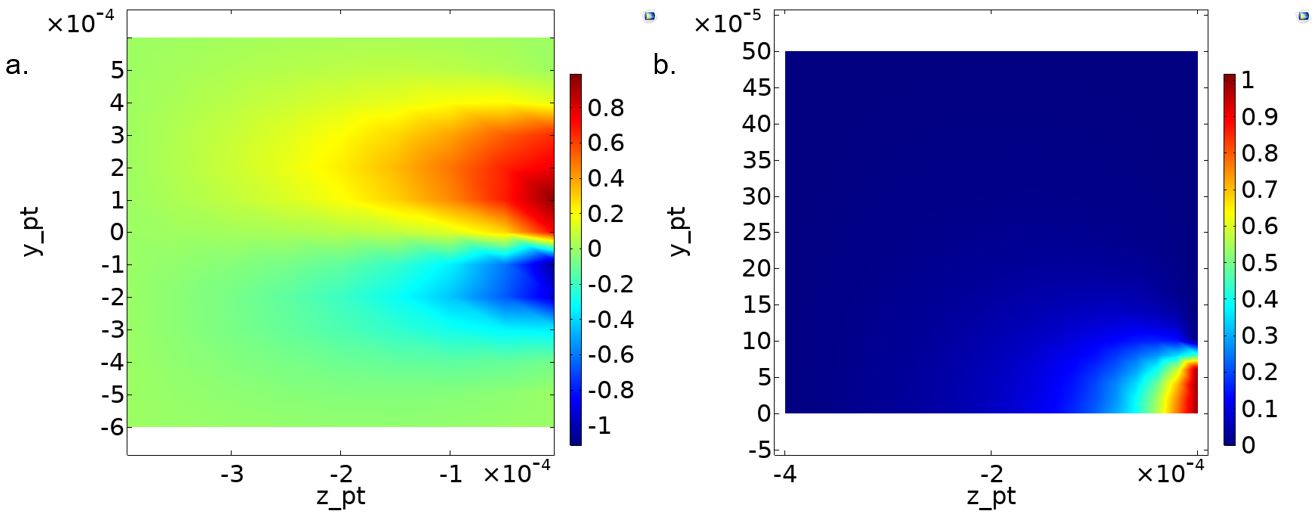} 
\caption{Charge sensitivity of the tantalum transmon.  \textbf{a.}  Shown is a COMSOL electromagnetic simulation of the charge sensitive volume surrounding the tantalum transmon.  The z-direction points into the substrate and the y-direction points from one transmon paddle to the other.  The device is centered on the top of the substrate at $x,y$ = 0,0.  Units of length on the axes are in meters.  The simulation was performed by placing an electron charge at each lattice point in the substrate and then calculating the induced charge on the device.  The amount of charge induced on a paddle is shown by the color bar.  \textbf{b.}  Shown is the analogous calculation performed for a single-island transmon with dimensions corresponding to the device measured in \cite{wilen20}.  The cylindrical symmetry of that transmon was utilized to reduce calculation time.  As before, the device is centered at $x,y$ = 0,0.  The differential design of the tantalum transmon is employed to reduce the charge sensitive volume but, owing to the much greater size of the tantalum device, still has a greater charge sensitive volume than the single-island transmon used in \cite{wilen20}.}
\end{figure*}

\nocite{*}

\bibliography{zSM_TaLFN_v2}